\documentclass[a4paper, 12pt, openany, oneside]{article}
\usepackage[cp1251]{inputenc} 
\usepackage[english, russian]{babel} 
\usepackage{amsmath, latexsym, amssymb, array, graphics, 
amsfonts, amsmath, bm}
\makeatletter

\renewcommand{\Re}{\mathop{\rm Re\,}}
\renewcommand{\Im}{\mathop{\rm Im\,}}

\makeatother
\usepackage{indentfirst}
\renewcommand{\baselinestretch}{1.2}
\usepackage[dvips]{graphicx}
\usepackage[flushleft,small]{caption2}

\begin{document}
\renewcommand{\abstractname}{ Abstract}
\renewcommand{\contentsname}{Contents}
\thispagestyle{empty}
\large
\renewcommand{\refname}{\begin{center} REFERENCES\end{center}}

 \begin{center}
\bf Analytical solution of the second Stokes problem
with variable amplitude on behaviour
of gas over  oscillation surface. Part I: eigenvalues and eigensolutions
\end{center}\medskip
\begin{center}
  \bf A. V. Latyshev\footnote{$avlatyshev@mail.ru$},
  E. A. Bedrikova\footnote{$bedrikova@mail.ru$}
\end{center}\medskip

\begin{center}
{\it Faculty of Physics and Mathematics,\\ Moscow State Regional
University, 105005,\\ Moscow, Radio str., 10--A}
\end{center}\medskip

\tableofcontents
\setcounter{secnumdepth}{4}

\begin{abstract}
In the present work the second Stokes problem about
behaviour of the rarefied gas filling half-space  is formulated. A plane
limiting half-space makes harmonious fluctuations
with variable amplitude in the plane. The amplitude changes
on the exponential law.
The kinetic equation with
model integral of collisions in the form $\tau$-model is used.
The case of diffusion reflexions of gas molecules  from a wall is considered.

Eigen solutions (continuous modes) of the initial
kinetic equation corresponding  to the continuous spectrum are searched.
Properties of dispersion function are studied. It is investigated
the discrete spectrum of the problem consisting of zero of the dispersion
functions in the complex plane. It is shown, that number of zero
of dispersion function to equally doubled index of problem coefficient.
The problem coefficient is understood as the relation of boundary values
of dispersion function from above and from below on the real  axis.

Further are eigen solutions (discrete modes)
of the initial kinetic equation corresponding to the discrete spectrum are searched.
In the end of work the general solution of the kinetic equation in the form
expansion  under  eigen solutions with unknown coefficients corresponding
to discrete and continuous spectra  is constructed.

\begin{center}
  \bf Реферат
\end{center}

В настоящей работе сформулирована вторая задача Стокса о
поведении разреженного газа, заполняющего полупространство. Плоскость,
ограничивающая полупространство, совершает гармонические колебания
с переменной амплитудой в своей плоскости. Амплитуда изменяется
по экспоненциальному закону.

Используется кинетическое уравнение с
модельным интегралом столкновений в форме $\tau$--модели.
Рассматривается случай диффузного отражения молекул газа от стенки.

Находятся собственные решения (непрерывные моды) исходного
кинетического уравнения, отвечающие непрерывному спектру.
Изучаются свойства дисперсионной функции. Исследуется
дискретный спектр задачи, состоящий из нулей дисперсионной
функции в комплексной плоскости. Показано, что число нулей
дисперсионной функции равно удвоенному индексу коэффициента
задачи. Под коэффициентом задачи понимается отношение граничных значений
дисперсионной функции сверху и снизу на действительной оси.
Далее находятся собственные решения (дискретные моды)
исходного кинетического уравнения, отвечающие дискретному спектру.

В конце работы составляется общее решение кинетического
уравнения в виде
разложения по собственным решениям с неизвестными
коэффициентами, отвечающими дискретному и непрерывному спектрам.
\medskip

{\bf Key words:} the second Stokes problem, collisional gas,
eigenvalues, eigen\-func\-tions, characteristic equation,
dispersion function, expansion by eigen\-func\-tions.

\medskip

PACS numbers: 05.20.Dd Kinetic theory, 47.45.-n Rarefied gas dynamics,
02.30.Rz Integral equations, 51. Physics of gases, 51.10.+y Kinetic and
transport theory of gases.
\end{abstract}

\begin{center}
\item{} \section{\bf Introduction}
\end{center}

Задача о поведении газа над движущейся поверхностью в последние годы
привлекает пристальное внимание \cite{0} -- \cite{15}.
Это связано с развитием современных технологий, в частности,
технологий наноразмеров.
В \cite{Yakhot} -- \cite{15} эта задача решалась численными или
приближенными методами. В настоящей работе показано, что эта
задача допускает аналитическое решение. Аналитическое решение
строится с помощью теории обобщенных функций и сингулярных
интегральных уравнений.

Впервые задача о поведении газа над стенкой, колеблющейся в
своей плоскости, была рассмотрена Дж. Г. Стоксом \cite{Stokes}.
Задача решалась гидродинамическим методом без учёта эффекта
скольжения. Обычно такую задачу называют второй задачей
Стокса \cite{Yakhot}--\cite{SS-2002}.

В последние годы на тему этой задачи появился ряд публикаций.
В работе \cite{Yakhot} рассматривается бесконечная
колеблющаяся поверхность. Задача рассматривается для любых
частот колебания поверхности. Из кинетического уравнения БГК
получено уравнение типа гидродинамического. Рассматриваются
гидродинамические граничные условия. Вводится коэффициент,
связывающий скорость газа на поверхности со скоростью
поверхности. Изотермическое скольжение не учитывается.
Получен вид графика зависимости силы трения на поверхности
от частоты колебаний поверхности. Показано, что в случае
высокочастотных колебаний сила трения, действующая на
поверхность, не зависит от частоты.

В работе \cite{SK-2007} получены коэффициенты вязкостного и
теплового скольжения с использованием различных модельных
уравнений. Использованы как максвелловские граничные условия,
так и граничные условия Черчиньяни --- Лэмпис \cite{Lampis}.

Наиболее близкая к решенной в первой и второй главах
диссертации \cite{15} задача решена в статье \cite{10}:
рассматривается газовый поток над бесконечной пластиной,
совершающей гармонические колебания в собственной плоскости.
Найдена скорость газа над поверхностью и сила, действующая на
поверхность со стороны газа. Для случая низких частот задача
решена на основе уравнения Навье --- Стокса. Изотермическое
скольжение не учитывалось. Для произвольных скоростей колебаний
поверхности задача решена численными методами на основе
кинетического уравнения Больцмана с интегралом столкновений
в форме БГК (Бхатнагар, Гросс, Крук). При этом рассматривался
только случай чисто диффузного отражения молекул от поверхности.
Дано аналитическое решение для случая колебаний высокой частоты.
И в этом случае рассматривается только чисто диффузное
отражение молекул от поверхности. В конце второй главы
диссертации \cite{15} проведено сопоставление результатов,
полученных в статье \cite{10} c результатами, полученными в
диссертации \cite{15}.

Работа \cite{11} является экспериментальным исследованием.
Изучается поток газа, создаваемый механическим резонатором
при различных частотах колебания резонатора. Эксперименты
показывают, что при низких частотах колебаний резонатора,
действующая на него со стороны газа сила трения прямо
пропорциональна частоте колебания резонатора. При высоких
частотах колебания резонатора ($~10^8$ Гц) действующая на
него сила трения от частоты колебаний не зависит.

В последнее время задача о колебаниях плоской поверхности в
собственной плоскости изучается и для случая неньютоновских
жидкостей \cite{5} и
\cite{6}.

В статье \cite{12} рассматривается пример практического
применения колебательной системы, подобной рассматриваемой во
второй задаче Стокса, в области нанотехнологий.

Общим существенным недостатком всех упомянутых теоретических
работ по решению второй задачи Стокса является отсутствие
учёта характера взаимодействия с поверхностью, т.е.
рассматривается только случай полной аккомодации
тангенциального импульса.

Коэффициент аккомодации тангенциального импульса является
величиной, зависящей от состояния поверхности. И если в
"естественном"\, состоянии значение этой величины как правило
близко к единице, то при специальной обработке поверхности
её значение можно уменьшить многократно \cite{13}, а значит
и существенно изменить характер взаимодействия поверхности
с прилегающим газом.

В условиях стремительного развития вакуумных технологий и
нанотехнологий, совершенствования авиационной и космической
техники весьма актуальным и целесообразным является
развитие направления исследований, связанного с определением
влияния взаимодействия молекул с поверхностью на перенос
импульса в системе "газ -- твёрдое тело"\, при произвольном
разрежении газа и установлением связи физических свойств
межфазной границы с макроскопическими газодинамическими
параметрами.

В диссертации \cite{15} были предложены два решения второй
задачи Стокса, учитывающие весь возможный диапазон
коэффициента аккомодации тангенциального импульса.
Эти решения отвечают соответственно гидродинамическому и
кинетическому описанию поведения газа над колеблющейся
поверхностью в режиме со скольжением.

Настоящая работа является продолжением работы \cite{0} и \cite{1}. В \cite{0}
вторая задача Стокса была решена аналитически для случая
постоянной амплитуды колебаний ограничивающей газ плоскости.
Настоящая работа является первой работой в серии работ авторов,
в которой мы намерены дать полное аналитическое решение
второй задачи Стокса с
переменной амплитудой. Амплитуда изменяется во времени по
экспоненциальному закону.

Для аналитического решения второй задачи Стокса используются сингулярные
интегральные уравнения с ядром Коши и обобщенные функции.
Настоящая работа --- первая из этой серии.

В п. 2 настоящей работы рассматривается постановка второй
задачи Стокса. Задача формулируется в общей постановке ---
с использованием граничных условий Максвелла (зеркально --
диффузных граничных условий). Далее задача будет рассматриваться
только для диффузных граничных условий. В качестве
кинетического уравнения рассматривается линеаризованное
кинетическое уравнение относительно функции $h(x_1,\mu)$.
Функция $h(x_1,\mu)$ (называемая также функцией распределения)
связана с полной функцией распределения соотношением
$$
f(\mathbf{r},\mathbf{v},t)=f_M(v)
[1+\sqrt{\beta}v_y e^{-i\omega t}h(\mathbf{r},\mathbf{v})],
$$
где
$$
f_M(v)=n\Big(\dfrac{m}{2\pi kT}\Big)^{3/2}\exp\Big(-\dfrac{m}{2kT}\Big)
$$
-- абсолютный максвеллиан.

Это уравнение получается путем
линеаризации модельного кинетического уравнения Больцмана
и интегралом столкновений в форме релаксационной $\tau$--модели.
Это так называемое модельное кинетическое БГК (Бхатнагар,
Гросс, Крук) уравнение.

Пластина (плоскость), ограничивающая полупространство с
разреженным газом совершает колебательные движения вдоль
оси $y$ с экспоненциально зависящей от времени амплитудой.

В качестве граничных условий используются два
условия. Одно из них --- граничное условие вдали от стенки
--- требует исчезания функции $h(x_1,\mu)$ вдали от стенки.
Второе условие --- условие на стенке --- вытекает из
требования диффузного отражения молекул от стенки.

Требуется
определить функцию распределения газовых молекул, найти
скорость газа в полупространстве и непосредственно у стенки,
найти силу трения, действующую со стороны газа на пластину,
найти мощность диссипации энергии пластины.

В п. 3 кинетическое уравнение упрощается путем представления
функции распределения в виде произведения $y$--компоненты
скорости молекул газа на новую неизвестную функцию. При этом
получается двухпараметрическое семейство кинетических уравнений
с комплексным параметром. Параметром уравнений служит
безразмерная величина частоты колебаний пластины. Эта величина
$\Omega_1=\omega/\nu=\omega \tau$ равна комплексной частоте колебаний
пластины $\omega$, деленной на величину частоты $\eta$
столкновений молекул газа, $\tau=1/\nu$ -- время между
двумя последовательными столкновениями молекулы.

В п. 4 находятся собственные решения (непрерывные моды)
исходного кинетического уравнения, отвечающие непрерывному
спектру. Последний совпадает с действительной положительной
полуосью. Эти собственные решения находятся в пространстве
обобщенных функций (распределений). Приводятся формулы
Сохоцкого для дисперсионной функции задачи, являющейся
основной в построенной теории.
Дисперсионная функция является кусочно аналитической
функцией с действительной осью в качестве линии скачков (разрывов).

В п. 5 исследуется дискретный спектр задачи, состоящий из
нулей дисперсионной функции в комплексной плоскости.
Применяется принцип аргумента из теории функций комплексного
переменного. Показано, что число нулей дисперсионной функции
равно удвоенному индексу коэффициента задачи. Под
коэффициентом $G(\mu)$ задачи понимается отношение граничных значений
дисперсионной функции сверху и снизу на действительной оси:
$G(\mu)=\lambda^+(\mu)/\lambda^-(\mu)$.

Выясняется, что на плоскости комплексной частоты
$(\omega_1,\omega_2)$
существует такая область $D$, ограниченная замкнутой линии
критических частот
$
\omega_1^*=\omega_1^*(\omega_2),
$
где
$$
\omega_1^*(\omega_2)=\max\limits_{0<\mu<+\infty}
\sqrt{[\Im\lambda^+(\mu)]^2-
[\omega_2+\Re\lambda^+(\mu)]^2},
$$
что при $(\omega_1,\omega_2)\in D^+$ индекс коэффициента
задачи равен единице: $\varkappa(G)=1$, а при
$(\omega_1, \omega_1) \notin \bar D^+$ индекс коэффициента
задачи равен нулю: $\varkappa(G)=0$.

Таким образом,
если точка $(\omega_1,\omega_2)$ находится внутри области $D^+$, то
дисперсионная функция имеет два комплексно -- значных
нуля, отличающихся лишь знаками в силу четности
дисперсионной функции. Если точка параметров $(\omega_1,\omega_2)$
находится во внешности замкнутой области $D^+$, то индекс задачи
равен нулю, т.е. дисперсионная функция комплексно --
значных нулей не имеет.

Далее находятся собственные решения (дискретные моды)
исходного кинетического уравнения, отвечающие дискретному спектру.

В конце работы составляется общее решение кинетического
уравнения в виде суммы собственного дискретного решения,
умноженного на неизвестную постоянную, и интеграла по
непрерывному спектру от собственных решений, отвечающих
непрерывному спектру, умноженных на неизвестную функцию.
Эти неизвестные постоянная и функция называются
соответственно коэффициентами дискретного и непрерывного
спектров. Эти неизвестные находятся из граничных условий.
Этому вопросу будут посвящены следующие наши работы.

\begin{center}
 \item{}\section{Linear kinetic equation for gas oscillations problem}
\end{center}

Пусть разреженный одноатомный газ занимает полупространство $x>0$
над плоской твердой поверхностью, лежащей в плоскости $x=0$.
Поверхность $(y,z)$ совершает гармонические колебания вдоль оси $y$
по закону $u_s(t)=u_0 e^{-i\omega t}$, где $\omega=\omega_1+i\omega_2$.
Эта запись означает, что плоскость совершает гармонические
колебания с частотой $\omega_1$ и экспоненциально переменной
амплитудой $A(t)=u_0e^{\omega_2t}$. Таким образом, при $\omega_2<0$
колебания являются экспоненциально затухающими, а при
$\omega_2>0$ -- экспоненциально возрастающими.

Возьмем кинетическое уравнение релаксационногго типа с
интегралом столкновений БГК (Бхатнагар, Гросс и Крук)
$$
\dfrac{\partial f}{\partial t}+v_x\dfrac{\partial f}{\partial x}=
\dfrac{f_{eq}-f}{\tau},
$$
где  $\tau$ -- время между двумя последовательными столкновениями
молекул, $\nu=1/\tau$ -- частота столкновений газовых молекул,
$f_{eq}$ -- равновесная функция распределения,
$$
f_{eq}=n\Big(\dfrac{m}{2\pi k T}\Big)^{3/2}\exp
\Big(-\dfrac{m}{2kT}\Big[v_x^2+(v_y-u_y(t,x))^2+v_z^2\Big]\Big),
$$
где  $m$ -- масса молекулы, $k$ -- постоянная Больцмана, $T$ --
температура газа, $u_y(t,x)$ -- массовая скорость газа,
$$
u_y(t,x)=\dfrac{1}{n}\int v_yf(t,x,\mathbf{v})d^3v,
\eqno{(1.1)}
$$
$n$ -- числовая плотность (концентрация) газа.

Концентрация газа и его температура считаются постоянными
в линеаризованной постановке задачи.

Далее будем линеаризовать кинетическое уравнение и искать
функцию распределения в виде
$$
f(t,x,\mathbf{v})=f_M(v)(1+\varphi(t,x,\mathbf{v})),
$$
$f_M(v)$ -- абсолютный максвеллиан,
$$
f_M(v)=n\Big(\dfrac{m}{2\pi kT}\Big)^{3/2}\exp\Big(-\dfrac{mv^2}{2kT}\Big).
$$

Получим линеаризованное кинетическое уравнение
$$
\dfrac{\partial \varphi}{\partial t}+
v_x\dfrac{\partial \varphi}{\partial x}+\nu\varphi(t,x,\mathbf{v})=
\dfrac{\nu m}{kT}v_yu_y(t,x).
\eqno{(1.2)}
$$

Введем безразмерные скорости и параметры: безразмерную скорость молекул:
$$
\mathbf{C}=\sqrt{\beta}\mathbf{v}=\dfrac{\mathbf{v}}{v_T},\qquad \;\beta=\dfrac{m}{2kT},
$$
безразмерную массовую скорость $U_y(t,x)=\sqrt{\beta}u_y(t,x)$,
безразмерное время $t_1=\nu t$, безразмерную координату
$$
x_1=\nu\sqrt{\dfrac{m}{2kT}}x=\nu \sqrt{\beta}x=\dfrac{x}{v_T\tau}=\dfrac{x}{l},
$$
где $l=v_T\tau$ -- средняя длина свободного пробега газовых молекул, и
безразмерную скорость колебаний пластины $U_s(t)=U_0e^{-i\omega t}$,
где
$$
U_0=\sqrt{\beta}u_0=\dfrac{u_0}{v_T}
$$
-- безразмерная амплитуда скорости
колебаний границы полупространства в начальный момент времени,
$v_T=\dfrac{1}{\sqrt{\beta}}=\sqrt{\dfrac{2kT}{m}}$ -- тепловая скорость
движения молекул, имеющая порядок скорости звука.
Тогда уравнение (1.2)
может быть записано в виде:
$$
\dfrac{\partial \varphi}{\partial t_1}+
C_x\dfrac{\partial \varphi}{\partial x_1}+\varphi(t_1,x_1,\mathbf{C})=
{2C_y}U_y(t_1,x_1).
\eqno{(1.3)}
$$

Заметим, что для безразмерного времени $U_s(t_1)=U_0e^{-i\Omega
t_1}$, где безразмерная комплексная частота колебаний границы:
$$
\Omega=\dfrac{\omega}{\nu}=\dfrac{\omega_1+i \omega_2}{\nu}=
\Omega_1+i\Omega_2,
$$
где
$$
\Omega_1=\dfrac{\omega_1}{\nu}, \qquad \Omega_2=\dfrac{\omega_2}{\nu}.
$$

В задаче о колебаниях газа требуется найти функцию
распределения $f(t_1,x_1,\mathbf{C})$ газовых молекул.
Функция распределения связана, как уже указывалось,
с функцией $\varphi(t_1,x_1,C_x)$ соотношением:
$$
f(t_1,x_1,\mathbf{C})=f_M(C)\big[1+\varphi(t_1,x_1,C_x)\big],
\eqno{(1.4)}
$$
где
$$
f_M(C)=n\Big(\dfrac{\beta}{\pi}\Big)^{3/2}\exp(-C^2)
$$
-- есть абсолютный максвеллиан.

Затем на основании найденной функции распределения требуется
найти массовую скорость газа, значение
массовой скорости газа непосредственно у стенки. Кроме того,
требуется вычислить силу сопротивления газа, действующую на
колеблющуюся пластину, ограничивающую газ.

Подчеркнем, что задача о колебаниях газа решается в
линеаризованной постановке.
Линеаризация задачи проведена согласно (1.4) по безразмерной
массовой скорости $U_y(t_1,x_1)$ при условии, что
$|U_y(t_1,x_1)|\ll 1$. Это неравенство эквивалентно
неравенству $$|u_y(t_1,x_1)|\ll v_T,$$ где $v_T=1/\sqrt{\beta}$
-- тепловая скорость молекул, имеющая порядок скорости звука.

Величину безразмерной массовой скорости $U_y(t_1,x_1)$ согласно ее
определению (1.1):
$$
U_y(t_1,x_1)=\dfrac{1}{\pi^{3/2}}\int \exp(-C^2)C_y\varphi(t_1,x_1,
\mathbf{C})d^3C.
\eqno{(1.5)}
$$

С помощью (1.5) кинетическое линеаризованное уравнение (1.3)
записывается в виде:
$$
\dfrac{\partial \varphi}{\partial t_1}+
C_x\dfrac{\partial \varphi}{\partial x_1}+
\varphi(t_1,x_1,\mathbf{C}) =\dfrac{2C_y}{\pi^{3/2}}
\int\exp(-{C'}^2)C_y'\varphi(t_1, x_1,\mathbf{C'})\,d^3C'.
\eqno{(1.6)}
$$

Сформулируем зеркально--диффузные граничные условия. Будем
считать, что молекулы отражаются от стенки согласно
зеркально--диффузной схеме:
$$
f(t,x=0,\mathbf{v})=qf_0(v)+(1-q)f(t,x=0,\mathbf{v}),\qquad v_x>0,
$$
где $q$ -- доля молекул, отражающихся чисто диффузно, а $1-q$ --
доля молекул, отражающихся чисто зеркально.

Отметим, что при $q=0$ мы получаем чисто зеркальное отражение
молекул от стенки, а при $q=1$ -- чисто диффузное отражение.

Вдали от стенки будем предполагать, что молекулы имеют
максвелловское распределение по скоростям, т.е.
$$
f(t,x\to +\infty, \mathbf{v})=f_M(v).
$$

Теперь отсюда формулируем зеркально--диффузные граничные
условия, записанные относительно функции $\varphi(t_1,x_1,\mathbf{C})$:
$$
\varphi(t_1,0,\mathbf{C})=2qC_yU_s(t_1)+(1-q)\varphi(t_1,0,
-C_x,C_y,C_z),\quad C_x>0,
\eqno{(1.7)}
$$
и
$$
\varphi(t_1,x_1\to+\infty,\mathbf{C})=0.
\eqno{(1.8)}
$$

Итак, граничная задача о колебаниях газа сформулирована
полностью и состоит в решении уравнения (1.6) с граничными
условиями (1.7) и (1.8).

Отметим, что к выражению (1.5) для безразмерной массовой
скорости можно придти, исходя из определения размерной
массовой скорости газа (1.1). В самом деле, подставляя в
(1.1) выражение (1.4), приходим в точности к выражению (1.5).

\begin{center}
\item{}\section{\bf Decomposition of the boundary problem}
\end{center}

Учитывая, что колебания пластины рассматриваются вдоль оси
$y$, будем искать, следуя Черчиньяни \cite{16}, функцию
$\varphi(t_1,x_1,\mathbf{C})$ в виде
$$
\varphi(t_1,x_1,\mathbf{C})=C_yH(t_1,x_1,C_x).
\eqno{(2.1)}
$$
Тогда безразмерная массовая скорость (1.5) с помощью (2.1) равна
$$
U_y(t_1,x_1)=\dfrac{1}{2\sqrt{\pi}}\int\limits_{-\infty}^{\infty}
\exp(-C_x'^2)H(t_1,x_1,C_x')dC_x'.
\eqno{(2.2)}
$$

С помощью указанной выше подстановки (2.1) кинетическое
уравнение (1.6) преобразуется к виду:
$$
\dfrac{\partial H}{\partial t_1}+C_x\dfrac{\partial H}{\partial x_1}+
H(t_1,x_1,C_x)=\dfrac{1}{\sqrt{\pi}}\int\limits_{-\infty}^{\infty}
\exp(-C_x'^2)H(t_1,x_1,C_x')dC_x'.
\eqno{(2.3)}
$$

Граничные условия (1.7) и (1.8) преобразуются в следующие:
$$
H(t_1,0,C_x)=2qU_s(t_1)+(1-q)H(t_1,0,-C_x),\qquad C_x>0,
\eqno{(2.4)}
$$
$$
H(t_1,x_1\to +\infty, C_x)=0.
\eqno{(2.5)}
$$

Следующим шагом  выделим временную переменную, положив далее:
$$
H(t_1,x_1,C_x)=e^{-i\Omega t_1}h(x_1,C_x).
\eqno{(2.6)}
$$

Теперь вместо (2.3) мы получаем уравнение относительно
функции $h(x_1,C_x)$:
$$
C_x\dfrac{\partial h}{\partial x_1}+(1-i\Omega)h(x_1,C_x)
=\dfrac{1}{\sqrt{\pi}}\int\limits_{-\infty}^{\infty}
\exp(-C_x'^2)h(x_1,C_x')dC_x'.
\eqno{(2.7)}
$$

Граничные условия (2.4) и (2.5) переходят в следующие:
$$
h(0,C_x)=2qU_0+(1-q)h(0,-C_x),\qquad C_x>0,
\eqno{(2.8)}
$$
и
$$
h(x_1\to+\infty,C_x)=0.
\eqno{(2.9)}
$$

Тогда безразмерная массовая скорость согласно (2.2) и (2.6) равна:
$$
U_y(t_1,x_1)=\dfrac{e^{-i\Omega t_1}}{2\sqrt{\pi}}\int\limits_{-\infty}^{\infty}
\exp(-C_x'^2)h(x_1,C_x')dC_x'.
\eqno{(2.10)}
$$

Мы получили граничную задачу, состоящую в решении уравенния (2.7)
с граничными условиями (2.8) и (2.9). Скорость газа согласно
(2.10) будет вычислена во второй части нашей работы.

\begin{center}
\item{}\section{ \bf Eigen solutions of the continuous spectrum}
\end{center}

Перепишем граничную задачу (2.7), (2.8) и (2.9) в виде:
$$
\mu\dfrac{\partial h}{\partial x_1}+z_0h(x_1,\mu)=\dfrac{1}{\sqrt{\pi}}
\int\limits_{-\infty}^{\infty}\exp(-{\mu'}^2)h(x_1,\mu')d\mu',
\eqno{(3.1)}
$$
где
$$
z_0=1-i\Omega,
$$
и
$$
h(0,\mu)=2qU_0+(1-q)h(0,-\mu)d\mu,\qquad \mu>0,
\eqno{(3.2)}
$$
$$
h(+\infty,\mu)=0.
\eqno{(3.3)}
$$

Граничная задача (3.1)--(3.3) будет рассмотрена в следующей
части нашей работы.
Разделение переменных в уравнении (3.1) осуществляется
следующей подстановкой
$$
h_\eta(x_1,\mu)=\exp\Big(-\dfrac{x_1z_0}{\eta}\Big)\Phi(\eta,\mu),
\eqno{(3.4)}
$$
где $\eta$ -- параметр разделения, или спектральный параметр,
вообще говоря, комплексный.

Подставляя (3.4) в уравнение (3.1) получаем характеристическое
уравнение
$$
(\eta-\mu)\Phi(\eta,\mu)=\dfrac{\eta}{\sqrt{\pi}z_0}
\int\limits_{-\infty}^{\infty}
\exp(-{\mu'}^2)\Phi(\eta,\mu')d\mu'.
\eqno{(3.5)}
$$
Если ввести обозначение
$$
n(\eta)=\dfrac{1}{z_0}\int\limits_{-\infty}^{\infty}
\exp(-{\mu'}^2)\Phi(\eta,\mu')d\mu',
\eqno{(3.6)}
$$
то уравнение (3.5) может быть записано с помощью (3.6) в виде
$$
(\eta-\mu)\Phi(\eta,\mu)=\dfrac{1}{\sqrt{\pi}}\eta n(\eta),\qquad
\eta\in \mathbb{C}.
\eqno{(3.7)}
$$

Уравнение (3.7) является конечным (недифференциальным) уравнением.
Условие (3.6) называется нормировочным условием,
нормировочным интегралом, или просто нормировкой.

Решение характеристического уравнения для действительных значений
параметра $\eta$ будем искать в пространстве
обобщенных функций \cite{0}.
Обобщенное решение уравнения (3.7) имеет вид:
$$
\Phi(\eta,\mu)=\dfrac{1}{\sqrt{\pi}}\eta n(\eta)P\dfrac{1}{\eta-\mu}+
g(\eta)\delta(\eta-\mu),
\eqno{(3.8)}
$$
где $-\infty<\eta, \mu <+\infty$.

Здесь $g(\eta)$ -- произвольная непрерывная функция, определяемая из
условия нормировки, $\delta(x)$ -- дельта--функция Дирака, символ $Px^{-1}$
означает главное значение интеграла при интегрировании $x^{-1}$.
Подставляя (3.8) в (3.6), получаем уравнение, из которого находим
$$
n(\eta)\lambda(\eta)=\exp(-\eta^2)g(\eta),
$$
где $\lambda(z)$ -- дисперсионная функция, введенная равенством
$$
\lambda(z)=1-i\Omega+\dfrac{z}{\sqrt{\pi}}\int\limits_{-\infty}^{\infty}
\dfrac{\exp(-\tau^2)d\tau}{\tau-z}.
$$
Эту функцию можно преобразовать к виду:
$$
\lambda(z)=-i\Omega+\lambda_0(z),
$$
где $\lambda_0(z)$ -- известная функция из теории плазмы,
$$
\lambda_0(z)=\dfrac{1}{\sqrt{\pi}}\int\limits_{-\infty}^{\infty}
\dfrac{e^{-\tau^2}\tau d\tau}{\tau-z}.
$$
Собственные функции (3.8) определены с точностью
до мультипликативной "постоянной"\,$n(\eta)$:
$$
\Phi(\eta,\mu)=\Big[\dfrac{1}{\sqrt{\pi}}\eta P\dfrac{1}{\eta-\mu}+
\exp(\eta^2)\lambda(\eta)\delta(\eta-\mu)\Big]n(\eta).
\eqno{(3.9)}
$$

Собственные функции (3.9) называются собственными функциями не\-прерывного
спектра, ибо спектральный параметр $\eta$ непрерывным образом
заполняет всю действительную прямую.

Далее в силу однородности уравнения (3.1) можно считать, что
$$
n(\eta)\equiv 1.
$$

Таким образом, собственные решения уравнения (3.4) имеют вид
$$
h_\eta(x,\mu)=\exp\Big(-\dfrac{x_1}{\eta}z_0\Big)
\Big[\dfrac{1}{\sqrt{\pi}}\eta P\dfrac{1}{\eta-\mu}+
\exp(\eta^2)\lambda(\eta)\delta(\eta-\mu)\Big].
\eqno{(3.10)}
$$

Собственные решения (3.10) отвечают непрерывному спектру характеристического
уравнения, ибо спектральный параметр непрерывным образом
пробегает всю числовую прямую, т.е. непрерывный спектр $\sigma_c$
есть вся конечная часть числовой прямой: $\sigma_c=(-\infty,+\infty)$.

По условию задачи мы ищем решение, невозрастающее вдали от стенки.
Поэтому далее будем рассматривать положительную часть
непрерывного спектра. В этом случае собственные решения (3.10)
являются исчезающими вдали от стенки. В связи с этим спектром
граничной задачи будем называть положительную действительную
полуось параметра $\eta$:
$\sigma_c^{\rm problem}=(0,+\infty)$.

В заключение этого п. приведем формулы Сохоцкого для дисперсионной функции:
$$
\lambda^{\pm}(\mu)=\pm i\sqrt{\pi}\mu e^{-\mu^2}-i\Omega+
\dfrac{1}{\sqrt{\pi}}\int\limits_{-\infty}^{\infty}
\dfrac{e^{-\tau^2}\tau d\tau}{\tau-\mu}.
$$
Разность граничных значений дисперсионной функции отсюда равна:
$$
\lambda^+(\mu)-\lambda^-(\mu)=2\sqrt{\pi}\mu e^{-\mu^2}i,
$$
полусумма граничных значений равна:
$$
\dfrac{\lambda^+(\mu)+\lambda^-(\mu)}{2}=-i\Omega+\dfrac{1}{\sqrt{\pi}}
\int\limits_{-\infty}^{\infty}\dfrac{e^{-\tau^2}\tau d\tau}{\tau-\mu}.
$$

Выделим действительную и мнимую части граничных значений
дисперсионной функции на действительной оси:
$$
\lambda^{\pm}(\mu)=-i\Omega_1+\Omega_2\pm i \sqrt{\pi}\mu e^{-\mu^2}+
\lambda_0(\mu)=
$$
$$
=\Re\lambda^{\pm}(\mu)+i\Im\lambda^{\pm}(\mu),\qquad -\infty<\mu<+\infty,
$$
где
$$
\Re\lambda^{\pm}(\mu)=\Omega_2+\lambda_0(\mu), \qquad
\Im\lambda^{\pm}(\mu)=-\Omega_1\pm s(\mu),
$$
$$
s(\mu)=\sqrt{\pi}\mu e^{-\mu^2}, \qquad
\lambda_0(\mu)=\dfrac{1}{\sqrt{\pi}}\int\limits_{-\infty}^{\infty}
\dfrac{e^{-\tau^2}\tau d\tau}{\tau-\mu},
$$
а интеграл в выражении для $\lambda_0(\mu)$ понимается в
смысле главного значения.

Заметим, что на действительной оси действительная часть
дисперсионной функции $\lambda_0(\mu)$ имеет два нуля $\pm\mu_0$,
$\mu_0=0.924\cdots$. Эти два нуля в силу четности функции
$\lambda_0(\mu)$ различаются лишь знаками.

Отметим, что на действительной оси дисперсионную функцию
удобнее использовать в численных расчетах в одном из
следующих видах (см. \cite{19}):
$$
\lambda_0(\mu)=\dfrac{1}{\sqrt{\pi}}\int\limits_{-\infty}^{\infty}
\dfrac{e^{-\tau^2}\tau d\tau}{\tau-\mu}=\lim\limits_{\varepsilon\to 0}
\dfrac{1}{\sqrt{\pi}}\int\limits_{-\infty}^{\infty}
\dfrac{e^{-\tau^2}\tau (\tau-\mu) d\tau}{(\tau-\mu)^2+\varepsilon^2},
$$
$$
\lambda_0(\mu)=1-2\mu^2 \int\limits_{0}^{1}\exp(-\mu^2(1-t^2))dt,\qquad
\mu\in(-\infty,+\infty).
$$

\begin{center}
\item{}\section{\bf Eigen solutions of the discrete spectrum}
\end{center}

Разложим дисперсионную функцию в ряд Лорана по отрицательным степеням
переменного $z$ в окрестности бесконечно удаленной точки:
$$
\lambda(z)=-i\Omega-\dfrac{1}{2z^2}-\dfrac{3}{4z^4}-\dfrac{15}{8z^6}-\cdots,
\quad z\to \infty.
\eqno{(4.1)}
$$

Из разложения (4.1) видно, что при малых значениях $|\Omega|$
дисперсионная функция имеет два отличающиеся лишь знаками
комплексно--значных нуля:
$$
\pm\eta_0^{(0)}(\Omega)=\dfrac{1+i}{2\sqrt{\Omega}}.
$$

\begin{figure}[t]
\begin{center}
\includegraphics[width=17.0cm, height=8cm]{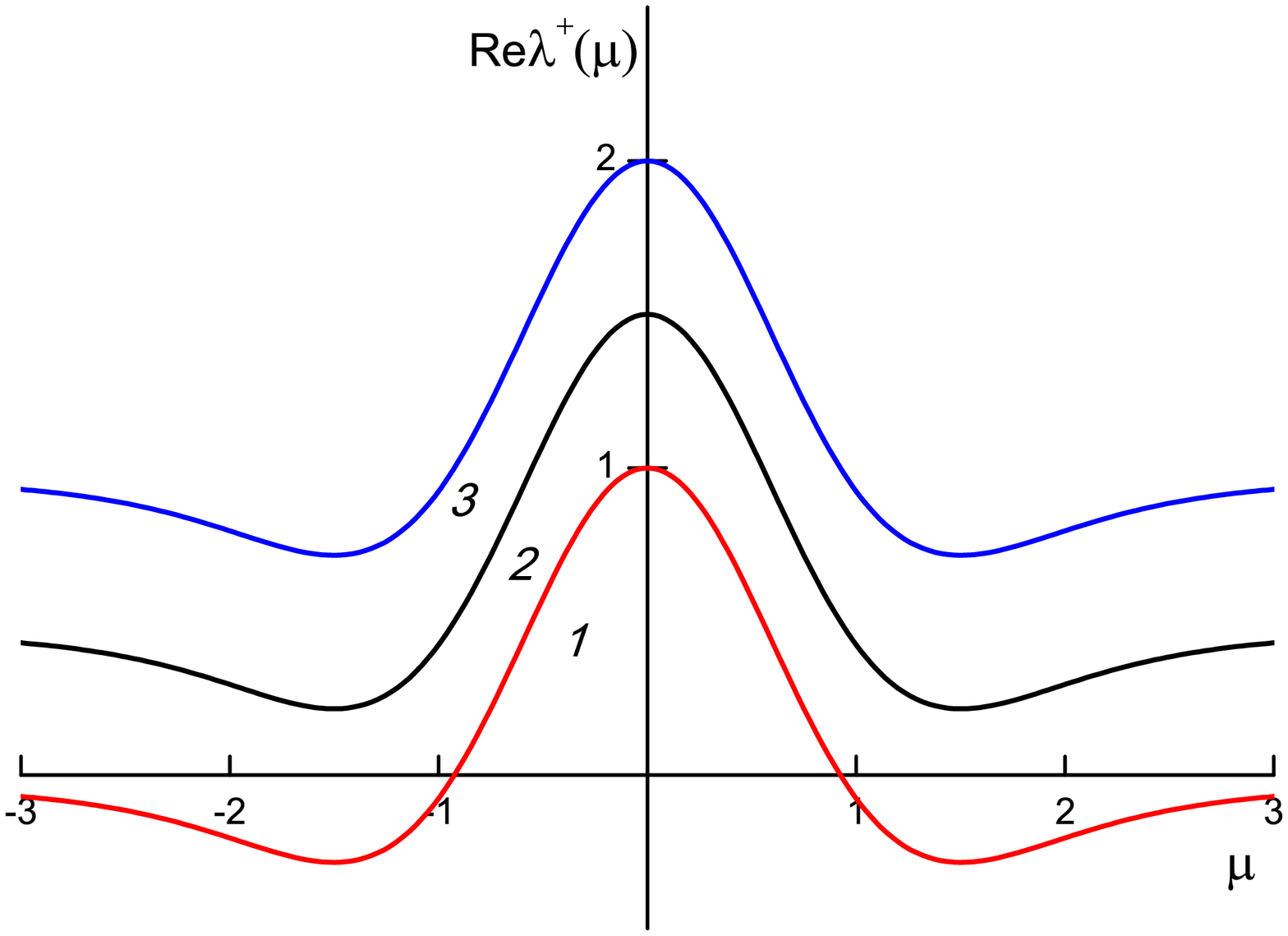}
\end{center}
\begin{center}
{{Рис 1. Действительная часть дисперсионной функции
$\Re\lambda^+(\mu)=\Omega_2+\lambda_0(\mu)$ на действительной оси.
 Кривые $1,2,3$ отвечают значениям
параметра $\Omega_2=0,0.5,1$}}
\end{center}
\end{figure}

\begin{figure}[h]
\begin{center}
\includegraphics[width=17.0cm, height=8cm]{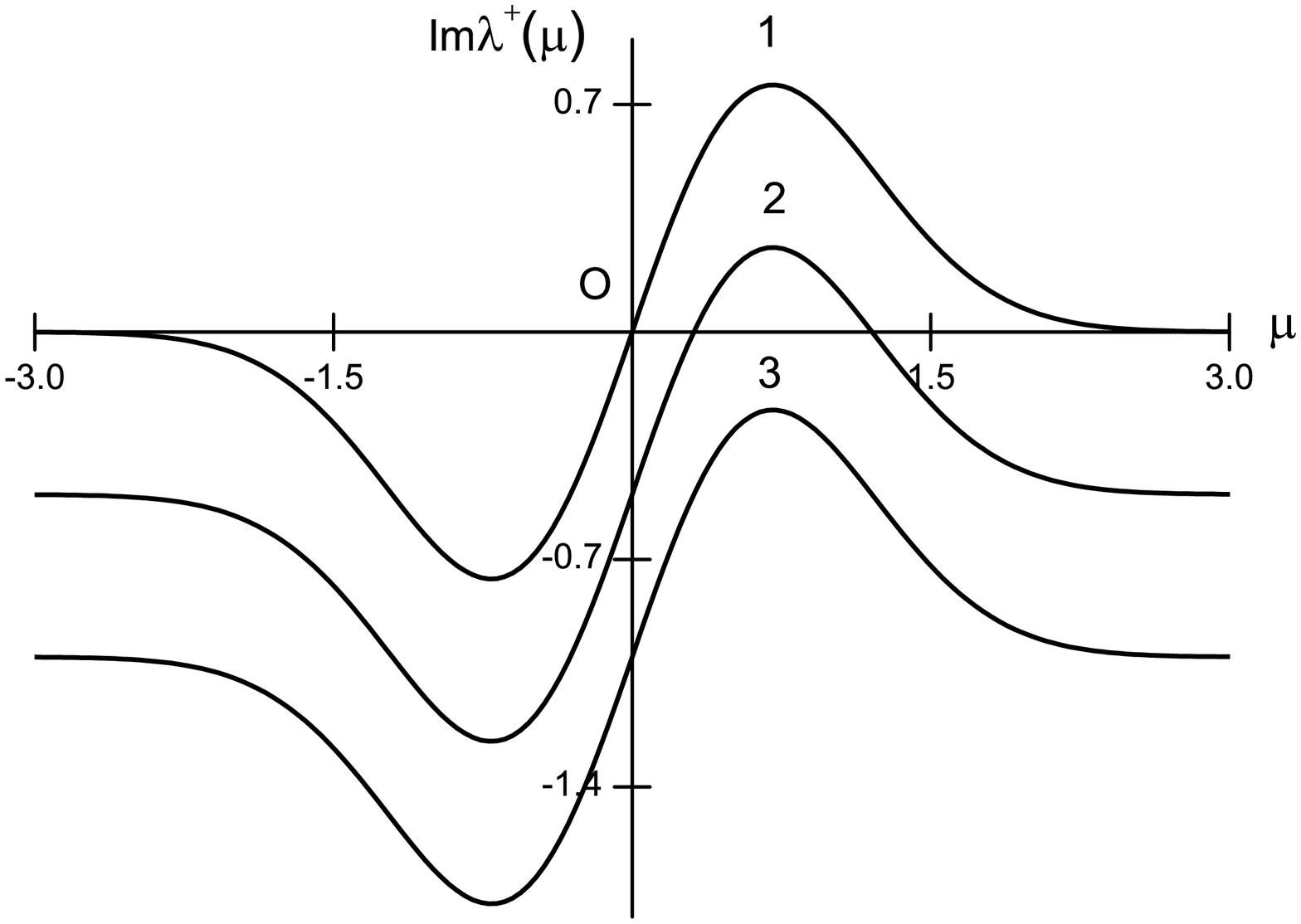}
\end{center}
\begin{center}
{{Рис 2. Мнимая часть дисперсионной функции $\Im\lambda^+(\mu)=-\Omega_1+
s(\mu)$ на действительной оси. Кривые $1,2,3$ отвечают значениям
параметра $\Omega_1=0,0.5,1$.}}
\end{center}
\end{figure}

Отсюда видно, что при $\Omega\to 0$ оба нуля дисперсионной функции
имеют пределом одну бесконечно удаленную точку $\eta_i=\infty$
кратности (порядка) два.

Из разложения (4.1) видно так же, что значение дисперсионной
функции в
бесконечно удаленной точки равно:
$$
\lambda(\infty)=-i\Omega.
$$

Применим теперь принцип аргумента для нахождения нулей дисперсионной
функции в верхней и нижней полуплоскостях. Этот подход является наиболее
общим.

Возьмем две прямые $\Gamma_\varepsilon^{\pm}$, параллельные
действительной оси и отстоящие от нее на расстоянии $\varepsilon,\;
\varepsilon>0$. Число $\varepsilon$ выберем настолько малым, чтобы
все нули дисперсионной функции лежали вне узкой полосы, заключенной
между прямыми $\Gamma_\varepsilon^{+}$ и $\Gamma_\varepsilon^{-}$.

Согласно принципу аргумента разность между числом нулей и числом
полюсов дисперсионной функции равно приращению ее логарифма:
$$
N-P=\dfrac{1}{2\pi i}\Bigg[\;\int\limits_{\Gamma_\varepsilon^+}+
\int\limits_{\Gamma_\varepsilon^-}\Bigg]\,d\,\ln \lambda(z).
\eqno{(4.2)}
$$

В (4.2) каждый нуль и полюс считаются столько раз, какова их
кратность, прямые $\Gamma_\varepsilon^{+}$ и
$\Gamma_\varepsilon^{-}$ проходятся соответственно в положительном и
отрицательном направлениях. Ясно, что дисперсионная функция полюсов не
имеет, т. е. $P=0$.

В пределе при $\varepsilon\to 0$ из равенства (4.2) получаем:
$$
N=\dfrac{1}{2\pi i}\int\limits_{-\infty}^{\infty}\,
d\ln \dfrac{\lambda^+(\mu)}{\lambda^-(\mu)}.
\eqno{(4.3)}
$$

Интеграл из (4.3) разобьем на два:
$$
\int\limits_{-\infty}^{\infty}
d\ln \dfrac{\lambda^+(\mu)}{\lambda^-(\mu)}=
\int\limits_{0}^{\infty}
d\ln \dfrac{\lambda^+(\mu)}{\lambda^-(\mu)}+
\int\limits_{-\infty}^{0}
d\ln \dfrac{\lambda^+(\mu)}{\lambda^-(\mu)}.
$$

Во втором интеграле сделаем замену переменной: $\tau\to -\tau$, при
такой замене имеем:
$$
\lambda^+(-\tau)=\lambda^-(\tau), \qquad
\lambda^-(-\tau)=\lambda^+(\tau).
$$
Следовательно, второй интеграл равен:
$$
\int\limits_{-\infty}^{0}
d\ln \dfrac{\lambda^+(\mu)}{\lambda^-(\mu)}=
-\int\limits_{0}^{\infty}
d\ln \dfrac{\lambda^+(-\mu)}{\lambda^-(-\mu)}.
$$
Таким образом,
$$
N=\dfrac{1}{2\pi i}\int\limits_{0}^{\infty}\,d\ln
\dfrac{\lambda^+(\mu)}{\lambda^-(\mu)}-\dfrac{1}{2\pi i}
\int\limits_{0}^{\infty}\,d\ln\dfrac{\lambda^+(-\mu)}{\lambda^-(-\mu)}=
$$
$$
=\dfrac{1}{2\pi i}\int\limits_{0}^{\infty}\,d\ln
\dfrac{\lambda^+(\mu)\lambda^-(-\mu)}{\lambda^-(\mu)\lambda^+(-\mu)}=
\dfrac{1}{2\pi i}\int\limits_{0}^{\infty}\,d\ln
\Bigg(\dfrac{\lambda^+(\mu)}{\lambda^-(\mu)}\Bigg)^2=
$$
$$
=\dfrac{1}{\pi i}\int\limits_{0}^{\infty}\,d\ln
\dfrac{\lambda^+(\mu)}{\lambda^-(\mu)}.
\eqno{(4.4)}
$$

Рассмотрим теперь на комплексной плоскости двухпараметрическое семейство кривых
$\Gamma=\Gamma(\Omega)$, заданных уравнением
\;$ z=G(\mu), \;0\leqslant \mu \leqslant +\infty$, где
$$
G(\mu)=\dfrac{\lambda^+(\mu)}{\lambda^-(\mu)}.
$$
Нетрудно проверить, что
$$
G(0)=1,\qquad \lim\limits_{\mu\to +\infty} G(\mu)=1.
$$

Эти равенства означают, что кривые $\Gamma(\Omega)$ являются
замкнутыми: они выходят из точки $z=1$ и заканчиваются в этой точке.
Согласно (4.4) имеем:
$$
N=\dfrac{1}{\pi i}\Big[\ln |G(\tau)|+i\arg G(\tau)\Big]
_0^{+\infty}=\dfrac{1}{\pi}\Big[\arg G(\tau)
\Big]_0^{+\infty}.
$$
Учитывая предыдущие равенства, отсюда получаем:
$$
N=\dfrac{1}{\pi}\Big[\arg G(t)\Big]_0^{+\infty}=2\varkappa(G),
\eqno{(4.5)}
$$
или
$$
N=2\varkappa(G),
$$
где $\varkappa=\varkappa(G)$ -- индекс функции $G(\mu)$ -- число
оборотов кривой $\Gamma(\Omega)$ относительно начала координат, совершаемых
в положительном направлении.

Из формулы (4.5) видно, что
$$
N=\dfrac{1}{\pi}\Big[\arg G(+\infty)-\arg G(0)\Big]=
\dfrac{1}{\pi}\arg G(+\infty),
\eqno{(4.6)}
$$
ибо $\arg G(0)=0$.

Введем угол $ \theta( \mu)= \arg G(\mu)$ --
главное значение аргумента функции $G(\mu)$, фиксированное
в нуле условием $ \theta(0)=0$.

Обозначим: $s(\mu)=\sqrt{\pi}\mu e^{-\mu^2}$.
Выделим действительную и мнимую части функции $G(\mu)$:
$$
G(\mu)=\dfrac{\Omega_2+\lambda_0(\mu)-i\Omega_1+is(\mu)}
{\Omega_2+\lambda_0(\mu)-i\Omega_1-is(\mu)}=
$$
$$
=\dfrac{(\Omega_2+\lambda_0(\mu))^2-s^2(\mu)+\Omega_1^2}
{(\Omega_2+\lambda_0(\mu))^2+[\Omega_1+s(\mu)]^2}+
i\dfrac{2(\Omega_2+\lambda_0(\mu))s(\mu)}
{(\Omega_2+\lambda_0(\mu))^2+[\Omega_1+s(\mu)]^2}.
$$

Отсюда видно, что
$$
\Re G(\mu)=\dfrac{(\Omega_2+\lambda_0(\mu))^2-s^2(\mu)+\Omega_1^2}
{(\Omega_2+\lambda_0(\mu))^2+[\Omega_1+s(\mu)]^2},
$$
$$
\Im G(\mu)=\dfrac{2(\Omega_2+\lambda_0(\mu))s(\mu)}
{(\Omega_2+\lambda_0(\mu))^2+[\Omega_1+s(\mu)]^2}.
\eqno{(4.7)}
$$

Таким образом, $G(\mu)=G_1(\mu)+iG_2(\mu)$, где
$$
G_1(\mu)=\Re G(\mu), \qquad G_2(\mu)=\Im G(\mu).
$$

Покажем, что на плоскости частот $(\Omega_1,\Omega_2)$
существует такая область $D^+$, что для точек
$(\Omega_1,\Omega_2)$, лежащих внутри этой области, индекс
задачи равен единице: $\varkappa(\Omega_1,\Omega_2)=1, (\Omega_1,\Omega_2)
\in D^+$, а для
точек $(\Omega_1,\Omega_2)$, принадлежащих ее внешности $D^-$,
индекс задачи равен нулю, т.е.
$\varkappa(\Omega_1,\Omega_2)=0, (\Omega_1,\Omega_2)
\in D^-$. Границу этой области $\partial D^+$назовем {\it линией критических
частот.}

Это означает, что если  $(\Omega_1,\Omega_2)\in D^+$,
то дисперсионная функция имеет два комплекснозначных нуля, а
если $(\Omega_1,\Omega_2)\in D^-$, то дисперсионная функция
нулей не имеет в верхней и нижней полуплоскостях.

Ясно, что если $G_1(\mu)$ или $G_2(\mu)$ не меняет своего знака,
то индекс задачи равен нулю. Из неравенства $G_2(\mu)\geqslant 0$
следует, что при $\Omega_2 \geqslant
\max\limits_{\mu}\{-\lambda_0(\mu)\}=0.284\cdots$, индекс задачи равен нулю.
Точно так же из неравенства $G_2(\mu)\le 0$ следует, что при
$\Omega_2 \leqslant -1$ индекс задачи равен нулю.

Теперь рассмотрим неравенство $G_1(\mu)\geqslant 0$. Из этого
неравенства следует, что если
$$
\Omega_1^2\geqslant \max\limits_{\mu}\{s^2(\mu)-[\Omega_2+
\lambda_0(\mu)]^2\},
$$
то индекс задачи равен нулю.

Образуем две линии предельных (или критических) частот
$|\Omega_1|=\Omega_1^*(\Omega_2)$, где
$$
\Omega_1^*(\Omega_2)=\max\limits_{\mu}\sqrt{ \{s^2(\mu)-[\Omega_2+
\lambda_0(\mu)]^2\}}.
$$

Введем область единичных частот (см. рис. 3)
$$
D^+=\{(\Omega_1,\Omega_2): |\Omega_1|< \Omega_1^*(\Omega_2),
-1<\Omega_2<\max\limits_{\mu}\{-\lambda_0(\mu)\}\}.
$$
Кривые $\Gamma(\Omega)$ согласно (4.7) определяются
параметрическими уравнениями
$$
\Gamma(\Omega): \quad x=\Re G(\mu),\quad y=\Im G(\mu),\quad
0 \leqslant \mu \leqslant +\infty.
\eqno{(4.8)}
$$

При $\Omega_1=\Omega_2=0$ мы имеем случай, рассмотренный в \cite{18}.

\begin{figure}[h]
\begin{center}
\includegraphics[width=17.0cm, height=8cm]{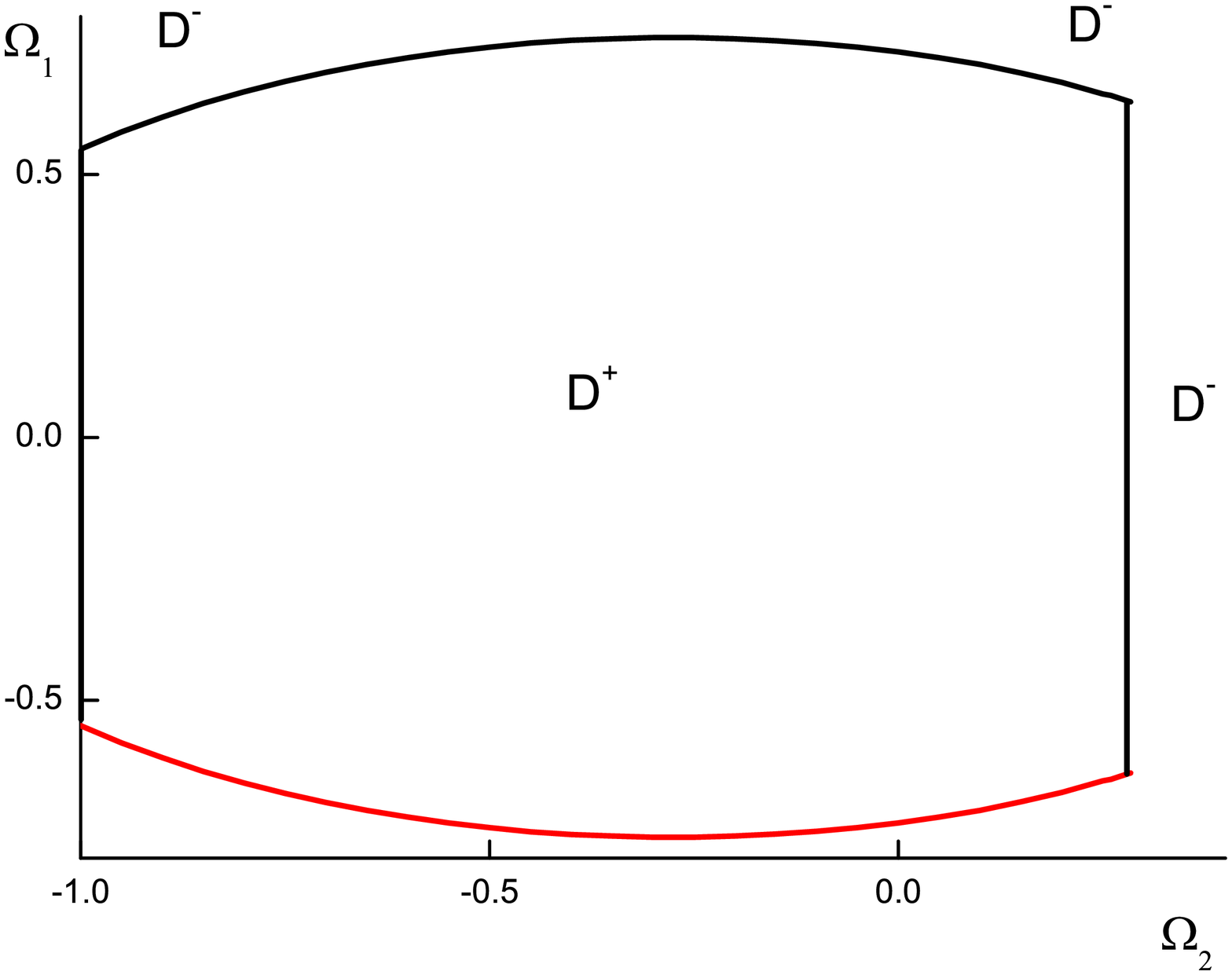}
\end{center}
\begin{center}
{{Рис 3. Области $D^+$ и $D^+$: при $(\Omega_1,\Omega_2)\in D^+$ индекс
$\varkappa=1$, $(\Omega_1,\Omega_2)\in D^-$ индекс
$\varkappa=0$.}}
\end{center}
\end{figure}
\begin{figure}[h]
\begin{center}
\includegraphics[width=17.0cm, height=6cm]{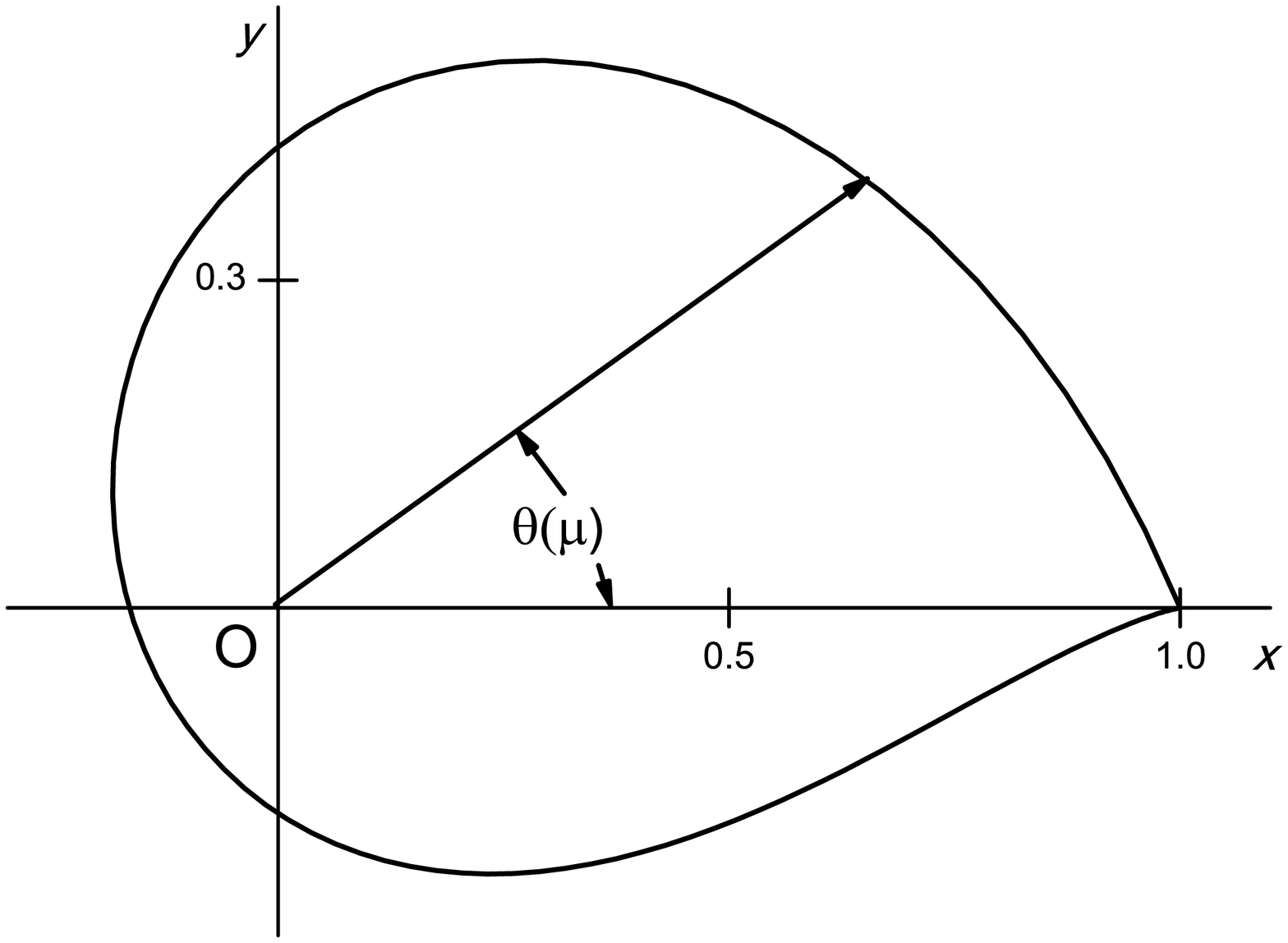}
{{Рис 4. Кривая $\Gamma(\Omega)$ является замкнутой и
охватывает начало координат при $(\Omega_1,\Omega_2)\in D^+$.
Индекс функции $G(\mu)$ равен единице, дисперсионная функция имеет
два комплексно--значных нуля.}}
\end{center}
\end{figure}

В этом случае кривая $\Gamma(0)$ охватывает один раз начало
координат. В самом деле, функция $\lambda_0(\mu)$ имеет
единственный нуль $\mu_0\approx 0.924$ на действительной положительной полуоси
оси, причем $\lambda_0(\mu)>0$ при $0\leqslant \mu<\mu_0$
и $\lambda_0(\mu)<0$ при $\mu_0<\mu<+\infty$.
Функция $y_1(\mu)=\lambda_0^2(\mu)-s^2(\mu)$ имеет два
нуля $\mu_1\approx 0.447$ и $\mu_2\approx 1.493$.
При этом $y_1(\mu)>0$ при $\mu\in [0,\mu_1)\cup (\mu_2,+\infty)$,
 а при $\mu\in (\mu_1,\mu_2)$: $y_1(\mu)<0$.
\begin{figure}[h]
\begin{center}
\includegraphics[width=17.0cm, height=6cm]{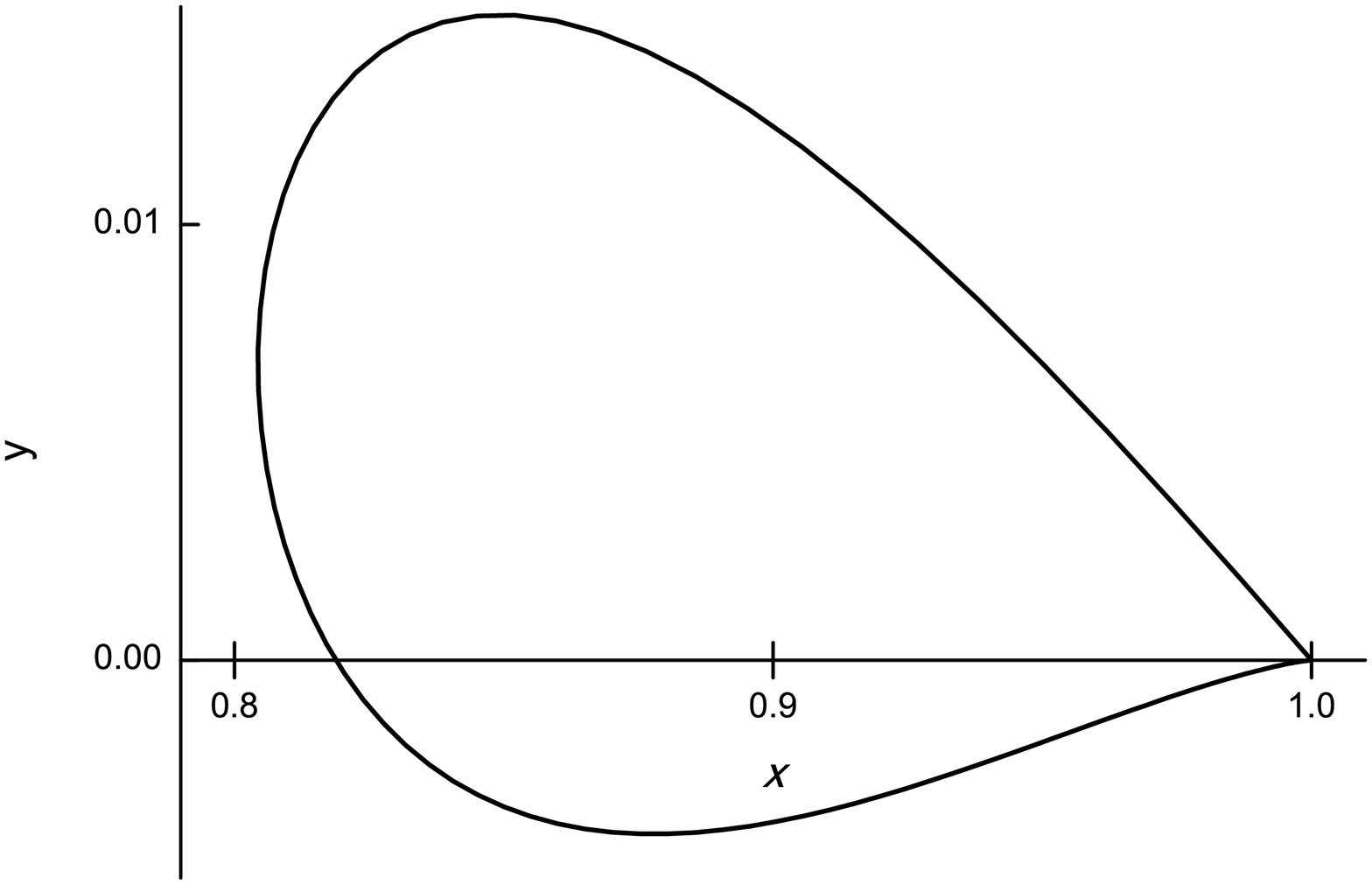}
\end{center}
\begin{center}
{{Рис 5. Кривая $\Gamma(\Omega)$ не охватывает начало
координат при $(\Omega_1,\Omega_2)\in D^-$. Индекс функции $G(\mu)$
равен нулю, дисперсионная функция имеет не имеет нулей в верхней
и нижней полуплоскостях.}}
\end{center}
\end{figure}
\begin{figure}[h]
\begin{center}
\includegraphics[width=17.0cm, height=8cm]{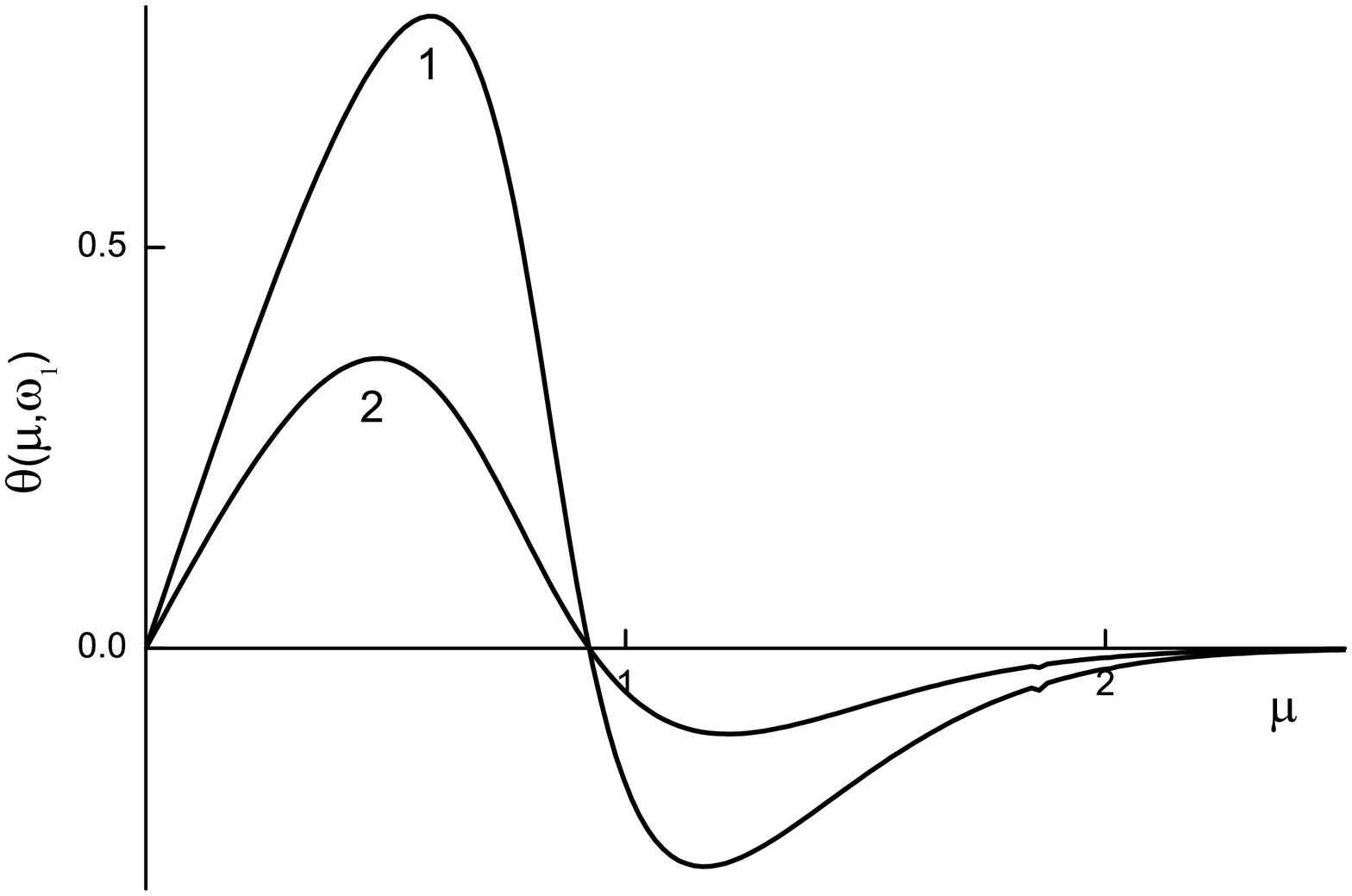}
\end{center}
\begin{center}
{{Рис 6. Зависимость угла $\theta=\theta(\mu,\Omega)$
от $\mu$ при различных значениях параметров $\Omega_1$ и $\Omega_2$
при $(\Omega_1,\Omega_2)\in D^-$. Индекс функции $G(\mu)$ равен
нулю. Приращение угла на полуоси равно нулю. Кривые $1$ и $2$
отвечает значениям $\Omega_1=1, \Omega_2=0$ и $\Omega_1=1.5, \Omega_2=0.$}}
\end{center}
\end{figure}
\clearpage

\begin{figure}[h]
\begin{center}
\includegraphics[width=17.0cm, height=8cm]{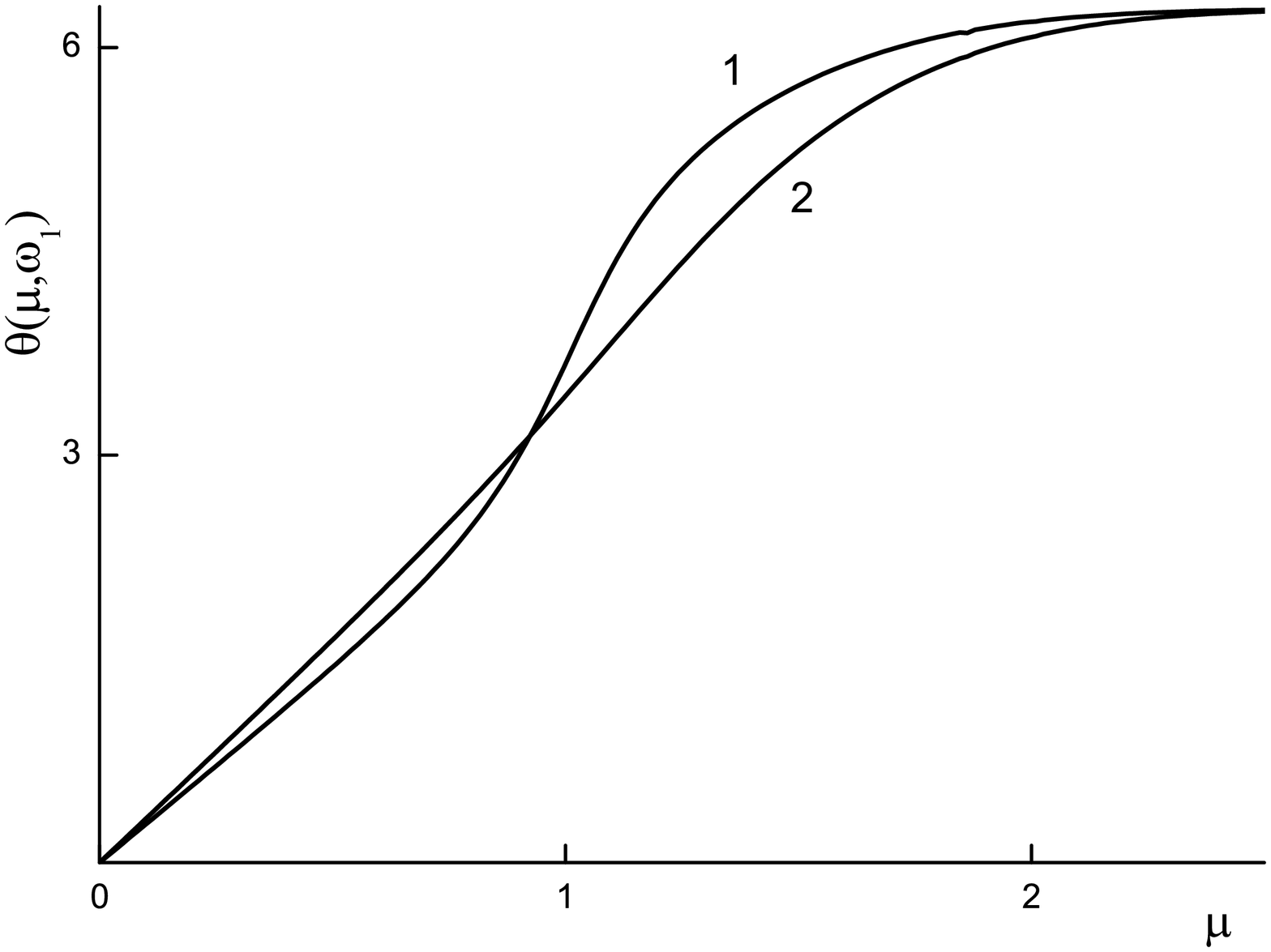}
\end{center}
\begin{center}
{{Рис 7. Зависимость угла $\theta=\theta(\mu,\Omega)$
от $\mu$ при различных значениях параметров $\Omega_1$  и
$\Omega_2$, при $(\Omega_1, \Omega_2)\in D^+$. Индекс функции $G(\mu)$
равен единице. Приращение угла на полуоси равно $2\pi$.
Кривые $1$ и $2$ отвечает значениям $\Omega_1=0.5, \Omega_2=0$ и
$\Omega_1=0.3, \Omega_2=0.$}}
\end{center}
\end{figure}

Теперь из соотношений (4.7) и (4.8) видно, что при
изменении $\mu$ от $0$ до $\mu_1$ кривая $\Gamma(0)$
выходит из точки $z=1$ и при $\mu=\mu_1$ оказывается в
точке на мнимой оси с координатой
$$
y(\mu_1)=\Im G(\mu_1)=\dfrac{2\lambda_0(\mu_1)s(\mu_1)}
{\lambda_0^2(\mu_1)+s^2(\mu_1)}>0.
$$

При этом кривая $\Gamma(0)$ описывает дугу, лежащую в
первой четверти. При изменении $\mu$ от $\mu_1$ до $\mu_0$
кривая описывает дугу, лежащую во второй четверти, и при
$\mu=\mu_0$ оказывается в точке
на действительной оси с координатой $x(\mu_0)=-1$.
При измененении $\mu$ от $\mu_0$ до $\mu_2$ кривая
$\Gamma(0)$ описывает дугу, лежащую в третьей четверти
и при $\mu=\mu_2$ оказывается в точке на мнимой оси с координатой
$$
y(\mu_2)=\dfrac{2\lambda_0(\mu_2)s(\mu_2)}{\lambda_0^2(\mu_2)+s^2(\mu_2)}<0,
\quad y(\mu_2)\geqslant -1.
$$

При дальнейшем изменении $\mu$ от $\mu_2$ до $+\infty$
кривая $\Gamma(0)$
лежит в четвертой четверти и заканчивается в точке $z=1$,
описывая один оборот вокруг начала координат.

Пусть теперь $\Omega_2=0$, а параметр $\Omega_1$ изменяется в пределах от
нуля до значения $\Omega_1^\circ=s(\mu_0)=
\sqrt{\pi}\mu_0e^{-\mu_0^2}\approx 0.697$.
Теперь корни $\mu_1$ и $\mu_2$ уравнения
$$
y_1(\mu,\Omega_1)=\lambda_0^2(\mu)-s^2(\mu)+\Omega_1^2
$$
становятся функциями параметра $\Omega_1$: $\mu_1=\mu_1(\Omega_1)$ и
$\mu_2=\mu_2(\Omega_1)$, причем $\mu_1(\Omega_1)< \mu_2(\Omega_1)$.

Нетрудно понять, что семейство кривых $\Gamma(\Omega_1)$
охватывает начало координат тогда и только тогда, когда
для нулей $\mu_1(\Omega_1), \mu_2$ и $\mu_2(\Omega_1)$
выполняется неравенство
$$
\mu_1(\Omega_1)<\mu_2<\mu_2(\Omega_1).
$$

Точка $\mu_2(\Omega_1)$ не зависит от $\Omega_1$, а при
возрастании $\Omega_1$ от $0$ до $\Omega_1^\circ$ точки
$\mu_1(\Omega_1)$ и $\mu_2(\Omega_1)$ сближаются навстречу
друг другу. При $\Omega_1=\Omega_1^\circ$ точки $\mu_0$ и
$\mu_2(\Omega_1)$ совпадают. Это означает, что кривая
$\Gamma(\Omega_1^\circ)$ проходит через начало координат.
Этому случаю можно приписать индекс $\varkappa(G)=1/2$.
\begin{figure}[h]
\begin{center}
\includegraphics[width=17.0cm, height=8cm]{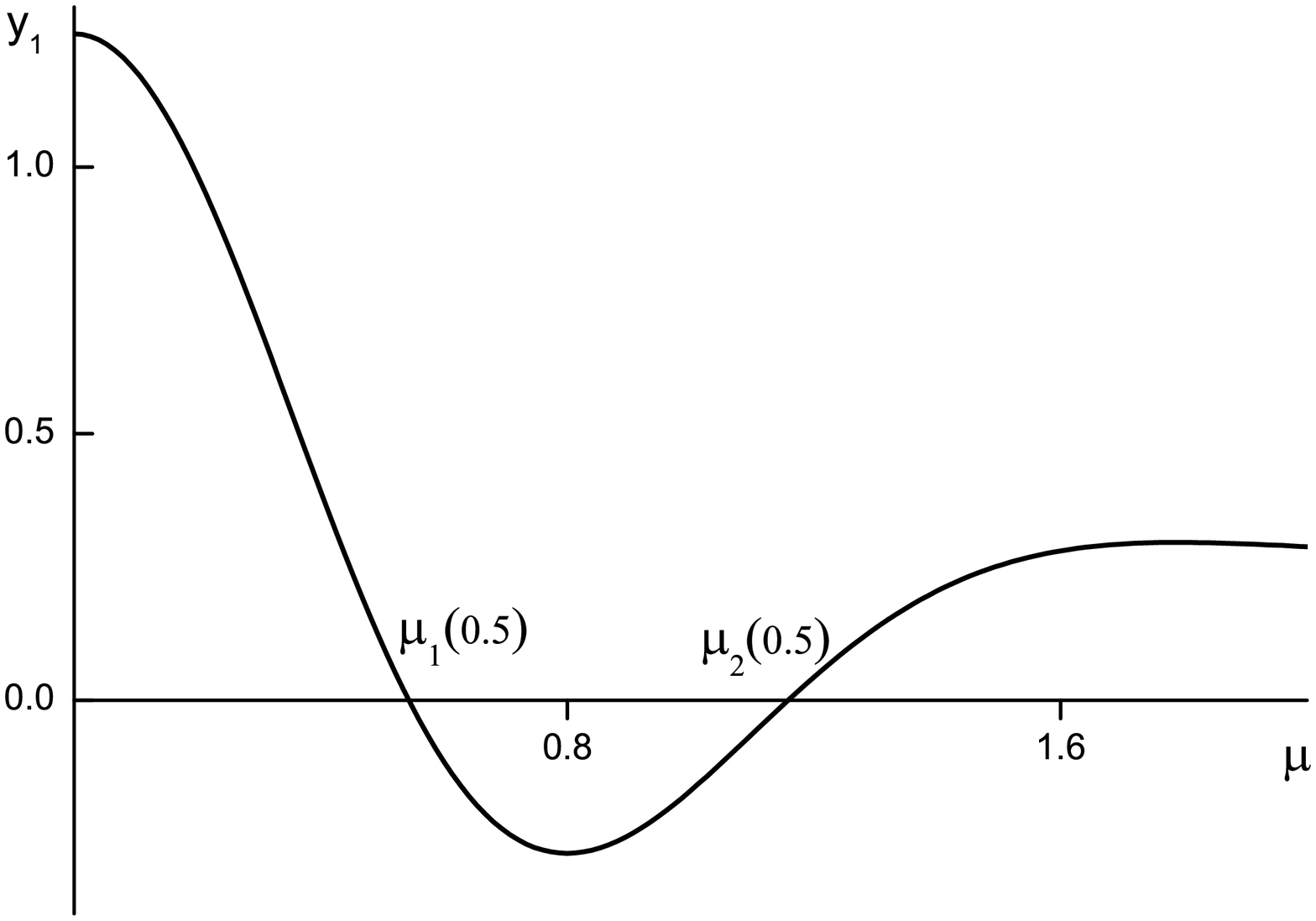}
\end{center}
\begin{center}
{{Рис 8. Случай $\Omega_2=0$. Нули $\mu_1(\Omega_1)$ и $\mu_2(\Omega_1)$
функции $y_1(\mu,\Omega_1)$ при $\Omega_1=0.5$; $\mu_1(0.5)=0.543,\quad
\mu_2(0.5)=1.158$.}}
\end{center}
\end{figure}

При дальнейшем возрастании $\Omega_1$ от $\Omega_1^\circ$
до $\Omega_1^*$ точка $\mu_2(\Omega_1)$ сначала перемещается
влево от точки $\mu_0$, не зависящей от $\Omega_1$, а далее
точки $\mu_1(\Omega_1)$ и $\mu_2(\Omega_1)$ совпадают:
$$
\mu_1(0.733)=\mu_2(0.733)=0.799.
$$

Затем при $\Omega_1 \geqslant \Omega_1^*$ функция $x(\mu)$
становится положительной при всех значениях $\mu, \;\mu \geqslant 0$.

Случай, когда $\Omega_2=0$, а параметр $\Omega_1$ изменяется в
пределах от нуля до значения $-\Omega_1^\circ$ рассматривается
полностью аналогично рассмотренному.

Общий случай, когда
$\Omega_2\in (-1,\max\limits_{\mu}\{-\lambda_0(\mu)\})$,
а $|\Omega_1|<\Omega_1^*(\Omega_2)$ более громоздок, но
рассматривается также аналогично предыдущему.

Итак, при $(\Omega_1,\Omega_2)\in D^+$ индекс равен
единице: $\varkappa(G)=1$, а при $(\Omega_1,\Omega_2)\in D^-$ индекс
равен нулю: $\varkappa(G)=0$.

Это означает, что при
$(\Omega_1,\Omega_2)\in D^+$ дисперсионная функция
имеет два нуля, а при $(\Omega_1,\Omega_2)\in D^-$ дисперсионная
функция нулей в верхней и нижней комлексной полуплоскости не имеет.

При $(\Omega_1,\Omega_2)\in D^+$ нули дисперсионной
функции обозначим через $\eta_0(\Omega)$ и $-\eta_0(\Omega)$.
В силу четности дисперсионной функции ее конечные нули
различаются только знаками, имея одинаковые модули.

Таким образом, дискретный спектр характеристического
уравнения, состоящий из нулей дисперсионной функции,
в случае $(\Omega_1,\Omega_2)\in D^+$ есть множество
из двух точек $\sigma_d(\Omega)=\{\eta_0(\Omega),
-\eta_0(\Omega)\}$.

При $(\Omega_1,\Omega_2)\in D^-$
дискретный спектр --- это пустое множество.

При $(\Omega_1,\Omega_2)\in D^+$ собственными
функциями характеристического уравнения являются следующие
два решения характеристического уравнения:
$$
\Phi(\pm \eta_0(\omega),\mu)=
\dfrac{1}{\sqrt{\pi}}\dfrac{\pm \eta_0(\Omega)}{\pm \eta_0(\Omega)-\mu}
$$
и два соответствующих собственных решения исходного
характеристического уравнения (3.1):
$$
h_{\pm \eta_0(\Omega)}(x_1,\mu)=
\exp \Big(-\dfrac{x_1z_0}{\pm \eta_0(\Omega)}\Big)
\dfrac{1}{\sqrt{\pi}}\dfrac{\pm \eta_0(\Omega)}
{\pm \eta_0(\Omega)-\mu}.
$$

Под $\eta_0(\Omega)$ будем понимать тот из нулей
дисперсионной функции, который обладает свойством:
$$
\Re \dfrac{1-i\Omega}{\eta_0(\Omega)}>0.
$$

Для этого нуля убывающее собственное решение кинетического
уравнения (3.1) имеет вид
$$
h_{\eta_0(\Omega)}(x_1,\mu)=\dfrac{1}{\sqrt{\pi}}
\exp\Big(-\dfrac{x_1z_0}{\eta_0(\Omega)}\Big)\dfrac{\eta_0(\Omega)}
{\eta_0(\Omega)-\mu}.
$$

Это означает, что дискретный спектр рассматриваемой
граничной задачи состоит из одной точки
$\sigma_d^{\rm problem}=\{\eta_0(\Omega)\}$
в случае $(\Omega_1,\Omega_2)\in D^+$.

При $\Omega\to 0$ оба нуля, как уже указывалось выше,
перемещаются в одну и ту же бесконечно удаленную точку.

Это значит, что в этом случае дискретный спектр
характеристического уравнения состоит из одной бесконечно
удаленной точки кратности два:
$\sigma_d(0)=\eta_i=\infty$ и является присоединенным к
непрерывному спектру. Этот спектр является также и
спектром рассматриваемой граничной задачи. Однако, в
этом случае дискретных (частных) решения ровно два:
$$
h_1(x_1,\mu)=1, \qquad h_2(x_1,\mu)=x_1-\mu.
$$

\begin{center}
  \item{}\section{General solution of the boundary problem}
\end{center}

Составим общее решение уравнения (3.1) в виде суммы
частного (дискретного) решения, убывающего вдали от стенки, и интеграла по
непрерывному спектру от собственных решений,
отвечающих непрерывному спектру:
$$
h(x_1,\mu)=\dfrac{a_0}{\eta_0-\mu}\exp\Big(-\dfrac{x_1z_0}
{\eta_0}\Big)
+\int\limits_{0}^{\infty}
\exp\Big(-\dfrac{x_1z_0}{\eta}\Big)\Phi(\eta,\mu)a(\eta)d\eta.
\eqno{(5.1)}
$$
Здесь $a_0$ -- неизвестный постоянный коэффициент,
называемый коэффициентом дискретного спектра, $a(\eta)$ --
неизвестная функция, называемая коэффициентом непрерывного
спектра, $\Phi(\eta,\mu)$ -- собственные функции
характеристического уравнения,
отвечающие непрерывному спектру и единичной нормировке.

Разложение (5.1) можно представить в явном виде:
$$
h(x_1,\mu)=\dfrac{a_0}{\eta_0-\mu}\exp\Big(-\dfrac{x_1z_0}
{\eta_0}\Big)+
$$
$$
+\int\limits_{0}^{\infty}
\exp\Big(-\dfrac{x_1}{\eta}z_0\Big)
\Big[\dfrac{1}{\sqrt{\pi}}\eta P\dfrac{1}{\eta-\mu}+
\exp(\eta^2)\lambda(\eta)\delta(\eta-\mu)\Big]a(\eta)d\eta.
\eqno{(5.2)}
$$
Функция $a(\eta)$ подлежит нахождению из граничных условий (3.2) и (3.3).

Разложение (5.2) можно представить в классическом виде:
$$
h(x_1,\mu)=\dfrac{a_0}{\eta_0-\mu}\exp\Big(-\dfrac{x_1z_0}
{\eta_0}\Big)+
$$
$$
+\dfrac{1}{\sqrt{\pi}}\int\limits_{0}^{\infty}
\exp\Big(-\dfrac{x_1z_0}{\eta}\Big)\dfrac{\eta a(\eta)d\eta}{\eta-\mu}+
\exp\Big(-\dfrac{x_1z_0}{\mu}+\mu^2\Big)\lambda(\mu)a(\mu)\theta_+(\mu),
\eqno{(5.3)}
$$
где $\theta_+(\mu)$ -- функция Хэвисайда,
$$
\theta_+(\mu)=\left\{\begin{array}{c}
                       1,\qquad \mu>0, \\
                       0,\qquad \mu<0
                     \end{array}.\right.
$$

\begin{center}
\item{}\section*{\bf Conclusions}
\end{center}

В настоящей работе сформулирована вторая задача
Стокса --- задача о поведении разреженного газа,
занимающего полупространство над стенкой, совершающей
гармонические колебания с переменной амплитудой,
экспоненциально зависящей от времени.

Рассматриваются зеркально--диффузные граничные условия.
Используется линеаризованное кинетическое уравнение,
полученное в результате линеаризации модельного
кинетического уравнения Больцмана в
релаксационном приближении.

Отыскивается структура дискретного и непрерывного
спектров задачи. Находятся собственные решения,
отвечающие дискретному и непрерывному спектрам.
Составляется общее решение кинетического уравнения
в виде разложения по собственным решениям.

\makeatother {\renewcommand{\baselinestretch}{1.2}

 \end{document}